\documentclass[twocolumn,aps,prd,amsmath,amssymb,superscriptaddress,showpacs]{revtex4-1}
\usepackage[utf8]{inputenc}
\usepackage{graphicx}
\usepackage{longtable}
\usepackage{amsbsy,amsmath}
\usepackage{hyperref}
\newcommand{\vect}[1]{\mathbf{#1}}
\newcommand{\gvect}{\boldsymbol}

\newcommand{\pdif}[2]{\frac{\partial{#1}}{\partial{#2}}}

\begin{document}

\title{Electrical Conductivity of Dense Quark Matter with Fluctuations and Magnetic Field Included}
\author{B.O. Kerbikov}
\email{borisk@itep.ru}

\author{M.A. Andreichikov}
\email{andreichicov@mail.ru}
\affiliation{Institute of Theoretical and Experimental Physics,\\ Bolshaya Cheremushhkinskaya 25, 117218, Moscow, Russia}
\affiliation{Moscow Institute of Physics and Technology,\\ Institutsky Pereulok 9, 141700, Dolgoprudny, Moscow Region, Russia}

\date{\today}

\begin{abstract}
  We investigate the electrical conductivity(EC) of dense quark matter in the vicinity of the phase transition line. We show that: (i) At high density Drude EC does not depend on the magnetic field up to $eB \sim 10^{19} \ G$. (ii) In the precritical region the fluctuation EC (paraconductivity) dominates over the Drude one.
\end{abstract}

\maketitle

\section{Introduction}
\label{sec-1}

QCD under extreme conditions has been a subject of the intense study for the last decade. A large body of experimental data on heavy-ion collisions obtained at RHIC and LHC has lead to a revolutionary change in our view on the properties of QCD matter at finite temperature and density. These properties depend on the location of the system in the QCD phase diagram, i.e., on the values of the temperature and the chemical potential. Roughly speaking, information obtained at RHIC and LHC corresponds to the high temperature and low density region. Our focus in the present paper is on the opposite regime of high density and moderate temperature. Such conditions may be realized in neutron stars and in future experiments at NICA and FAIR. On the theoretical side we understand much better what happens to quark-gluon matter at high $T$ and zero, or small $\mu$, than in the reverse situation. To a great extent this is due to the fact that zero $\mu$ and high $T$ region is accessible to Monte-Carlo simulations. According to the present understanding of the phase structure of the QCD matter attractive interaction between quarks in color antitriplet state leads at high density to the formation of the color superconducting phase \cite{1,2}. Some important features of this phase are, however, very different from that of a standard BCS superconductor \cite{3}. What is important for us here is that instead of an almost sharp dividing line between the normal and superconducting phases in the BCS case, in color superconductor the transition is significantly smeared and the effect of the precritical fluctuations is important. From the analysis below we infer the conclusion that fluctuation conductivity(or paraconductivity) \cite{4} is large and exceeds the Drude(or Boltzmann) one. Another question we address is the dependence of the EC on the magnetic field(MF). The outbreak of the interest to the behaviour of quark matter in strong MF is caused by the fact that MF of the order of $eB \sim \Lambda_{QCD}^2 \sim 10^{19} \ G$ ($GeV^2 \simeq 5.12 \cdot 10^{19} \ G$) is created(for a short time) in peripheral heavy-ion collisions at RHIC and LHC \cite{5,6}. The field of about 4 orders of magnitude less exists on the surface of magnetars, and may be of the order of $10^{17} \ G$ in its interior \cite{7}. A powerful way to investigate the nature of a certain substance is to study its response to the external perturbation, to the MF in our case. Below we show that due to a large value of the chemical potential in dense quark matter EC does not exhibit a drastic suppression with MF increasing. Such a behaviour is at odds with the typical behaviour of the Drude EC in condensed matter. 

The paper is structured as follows. In Section \ref{sec:2} we evaluate the Drude EC using the diagrammatic approach. Section \ref{sec:3} develops the Boltzmann equation for the EC. Here we show that Drude EC does not feel the influence of the MF until it reaches a huge value of about $10^{19} \ G$. In Section \ref{sec:4} we turn to the quantum contribution to the EC. We outline the general picture of the fluctuation phenomena in the precritical region. Then we introduce the fluctuation propagator and go on to compute the Aslamazov-Larkin paraconductivity. Our results are summarized in  Section \ref{sec:5}. Numerical estimates are presented here and the problems to be solved along the same lines are formulated.  

\section{Diagrammatic Computation}
\label{sec:2}

\begin{figure}
  \includegraphics[width=0.4\textwidth]{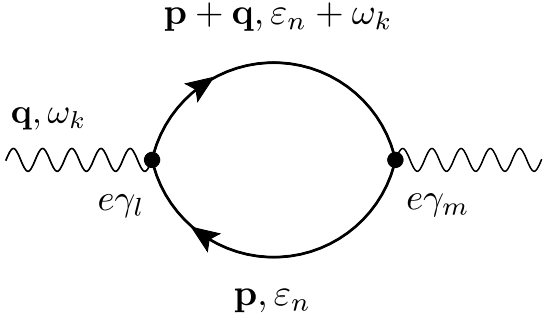}
  \caption{Diagrammatic representation of Drude EC}
  \label{fig:1}
\end{figure}

As a starting point we consider a diagrammatic derivation of the Drude EC of dense quark matter. The Drude EC can be derived from the Kubo two-point current-current correlator \cite{4,8}. The corresponding diagram is shown in Fig.\ref{fig:1}. One has
\begin{widetext}
\begin{equation}
  \label{eq:1}
  \sigma_{lm}(\vect{q}, \omega_k) = \frac{e^2T}{\omega_k} \sum_{\varepsilon_n} \int\frac{d\vect{p}}{(2\pi)^3} \mathrm{Tr} G(\vect{p}, \tilde{\varepsilon}_n) \gamma_l G(\vect{p} + \vect{q}, \tilde{\varepsilon}_n + \omega_k) \gamma_m 
\end{equation}
\end{widetext}

With u- and d-quarks included the electric charge is 
\begin{eqnarray}
  \nonumber
  \label{eq:2}
  e^2(N_f = 2) = C_{EM} \\
  = 4\pi \alpha \left[ \left(\frac{2}{3} \right)^2 + \left(\frac{1}{3} \right)^2 \right] = 0.051
\end{eqnarray}
Matsubara propagators entering in (\ref{eq:1}) have the form
\begin{equation}
  \label{eq:3}
  G(\vect{p}, \varepsilon_n) = \frac{1}{\gamma_0(i\tilde{\varepsilon}_n + \mu) - \gvect{\gamma}\vect{p} - m}
\end{equation}
Here $\omega_k = 2\pi Tk$, $\tilde{\varepsilon}_n = \varepsilon_n + \frac{1}{2\tau}\mathrm{sgn}\varepsilon_n$, $\varepsilon_n = \pi T(2n + 1)$, where $\mu$ is the chemical potential, $\tau$ is the momentum relaxation, or transport relaxation time. From the formal point of view $\tau$ regulates the pinch(collinear) singularities of(\ref{eq:1}) coming from the multiplication of the two propagators with momenta $(\vect{p}, \varepsilon_n)$ and $(\vect{p} + \vect{q}, \varepsilon_n + \omega_k)$ in the limit $|\vect{q}| \rightarrow  0$, $\omega_k \rightarrow 0$. From the physical point of view $\tau$ reflects the fact that quarks do not propagate through dense thermal medium without collisions. 

The loop diagram shown in Fig.\ref{fig:1} at finite $T$ and $\mu$ has been calculated in a number of papers, see, e.g., \cite{9,10}. We want to present the steps and assumptions which are distinctive for the physical situation of high density and moderate temperature. In order to calculate expression (\ref{eq:1}) we make use of the following conditions/assumptions:
\begin{equation}
  \label{eq:4}
  q \ll p
\end{equation}
\begin{equation}
  \label{eq:5}
  T \ll \mu
\end{equation}
Relation (\ref{eq:4}) is easily recognized as hard thermal loop approximation HTL \cite{11}. Indeed, the external momentum $q$ is assumed to be soft, but the internal momentum $p$ is hard. However, in our case this is not due to the high temperature as in standard HTL, but due to the fact that the dominant contribution to the integral in (\ref{eq:1}) comes from the vicinity of the Fermi surface with high $\mu$. Keeping in mind the above difference between the high temperature and high density regimes, we can follow the derivation presented in \cite{10} and obtain 
\begin{eqnarray}
  \nonumber
  \label{eq:6}
  \sigma(\vect{q}, \omega_k) = \sigma_{ll}(\vect{q}, \omega_k) \\
  = \frac{1}{3}e^2 \vect{v}^2 \tau \nu \int \frac{d\Omega}{4\pi} \frac{1}{1 + |\omega_k|\tau + i\vect{q}\vect{v}\tau}
\end{eqnarray}
Velocity $\vect{v}$ entering in (\ref{eq:6}) is the relativistic quark velocity near the Fermi surface $\vect{v}^2 = p_F^2/\mu^2 \rightarrow 1$ in the limit $m \rightarrow 0$. The quantity $\nu$ is the density of states near the Fermi surface. Under the condition $T \ll \mu$ it reads
\begin{equation}
  \label{eq:7}
  \nu = \frac{\mu p_F}{\pi^2}\left[ 1 + \frac{1}{3}\left(\frac{\pi T}{\mu} \right)^2 \right]
\end{equation}
Terms of the order $\left(T/\mu \right)^4$ and higher are omitted \cite{12,13}. It is instructive to compare $\nu$ given by (\ref{eq:7}) with the thermal asymptotic mass \cite{10}
\begin{equation}
  \label{eq:8}
  M^2 = \frac{T^2}{3} + \frac{\mu^2}{\pi^2}
\end{equation}
Next we perform the calculation of the integral (\ref{eq:6}). Exact calculation of  (\ref{eq:6}) yields
\begin{equation}
  \label{eq:9}
  \sigma(\vect{q}, \omega_k) = \frac{1}{6}e^2 \frac{v \nu}{iq} \ln \frac{1 + |\omega_k|\tau + i\vect{q}\vect{v}\tau}{1 + |\omega_k|\tau - i\vect{q}\vect{v}\tau}
\end{equation}
This general expression is most adequate for the ballistic regime when $|\omega_k|\tau \gg 1$, $qv\tau \gg 1$. Dense quark matter near the transition line is a highly disordered medium, the quark mean free path does not exceed 1 fm(see below). We shall therefore obtain from (\ref{eq:6}) an approximate expression relevant for the ``dirty'' regime $|\omega_k|\tau \ll 1$, $qv\tau \leq 1$. Expanding the integrand in (\ref{eq:6}) over these two parameters and integrating, we get
\begin{equation}
  \label{eq:10}
  \sigma(\vect{q}, \omega_k) = \frac{1}{3}e^2v^2\nu \frac{\tau}{1 + \omega \tau + \mathcal{D}\vect{q}^2 \tau}
\end{equation}
where $\omega = i\omega_k$, $\mathcal{D} = \frac{1}{3}v^2\tau^2$. This is a classical expression for Drude EC in the diffusion regime.

At this point let us remind the assumptions behind this result and indicate the imprints of the high density regime. The polarization operator in (\ref{eq:1}) has been calculated under the assumption of a soft external momentum $q$ and a hard internal one $p$. Physically this is tantamount to the proximity of $p$ to the Fermi surface. The above approximation can be called the Hard Density Loop in analogy with HTL. Another approximation is the expansion over $T/\mu$ in the expression (\ref{eq:7}) for the density of states. Finally we note that our result (\ref{eq:10}) differs from the nonrelativistic one only kinematically, namely in (\ref{eq:10}) $v = p_F/\mu$ and $\nu \sim \mu p_F$ instead of $v = p_F/m$ and $\nu \sim m p_F$ in the nonrelativistic case.

The next task is to investigate the influence of MF on EC. We shall be interested in the role played by the high density. The problem can be investigated either diagrammatically, or using the Boltzmann kinetic equation. We shall use the second approach which is technically simpler. 

\section{Boltzmann Equation Analysis}
\label{sec:3}

The Boltzmann equation in the relaxation time approximation reads \cite{14}
\begin{equation}
  \label{eq:11}
  -e(\boldsymbol{\mathcal{E}} + \vect{v} \times \vect{B})\pdif{f}{\vect{p}} = - \frac{f - f_0}{\tau}
\end{equation}
Here $\tau$ is the relaxation time, $f_0$ is the distribution function in the equilibrium
\begin{equation}
  \label{eq:12}
  f_0(p) = \frac{1}{\exp\left[\beta(E - \mu) \right] + 1}
\end{equation}
where $E^2 = \vect{p}^2 + m^2$. Equation (\ref{eq:11}) is written under the assumption of small inhomogeneity in space and time(i.e., we shall evaluate EC at $\omega_k = |\vect{q}| = 0$). For simplicity we consider the equation for the quark distribution neglecting antiquarks. At high density this is a trustworthy approximation which can be easily lifted \cite{15}. The solution of the kinetic equation (\ref{eq:11}) and the derivation of EC are described in textbooks \cite{13,14}. Our attention will be focused on the points reflecting the regime of high density and moderate temperature. It will be shown that the role of the mass of the electric charge carrier will be taken by the chemical potential $\mu$, and since $\mu \gg m$ the damping of EC by MF is drastically reduced. 

Following \cite{13} we work in the linear in $\boldsymbol{\mathcal{E}}$ approximation and substitute $f$ by $f_0$ in the term proportional to $\boldsymbol{\mathcal{E}}$ in (\ref{eq:11}). The distribution $f_0$ is isotropic, therefore
\begin{equation}
  \label{eq:13}
  (\vect{v} \times \vect{B}) \pdif{f_0}{\vect{p}} = (\vect{v} \times \vect{B})\vect{v} \pdif{f_0}{E} = 0
\end{equation}
where $\vect{v} = \vect{p}/E \simeq \vect{p}/\mu$ in the vicinity of the Fermi surface. This is where $\mu$ comes into play. Then (\ref{eq:11}) takes form
\begin{equation}
  \label{eq:14}
  -e\vect{v}\boldsymbol{\mathcal{E}}\pdif{f_0}{E} - e(\vect{v} \times \vect{B})\pdif{(f - f_0)}{\vect{p}} = -\frac{f - f_0}{\tau}
\end{equation}
The solution of (\ref{eq:14}) is parametrized as 
\begin{equation}
  \label{eq:15}
  f = f_0 - \vect{v}\vect{C}(\varepsilon)\pdif{f_0}{E}
\end{equation}
This ansatz (\ref{eq:14}) transforms into
\begin{equation}
  \label{eq:16}
  -e\boldsymbol{\mathcal{E}} + (\gvect{\Omega} \times \vect{C}) = \frac{\vect{C}}{\tau}
\end{equation}
where $\gvect{\Omega} = e\vect{B}/\mu$. Solution of this equation reads 
\begin{eqnarray}
  \nonumber
  \label{eq:17}
  f = f_0 + \frac{e\tau}{1 + \Omega^2\tau^2} \left[ \vect{v} + \tau^2(\vect{v}\gvect{\Omega})\gvect{\Omega} \right.\\
  \left. + \tau(\vect{v} \times \gvect{\Omega}) \right]\boldsymbol{\mathcal{E}} \left( \pdif{f_0}{E} \right)
\end{eqnarray}
Electric current is given by 
\begin{eqnarray}
  \nonumber
  \label{eq:18}
  \vect{j} = 2e^2 \int d\vect{p} \left(-\pdif{f_0}{E} \right) \frac{\tau}{1 + \Omega^2\tau^2}\\
 \times \left[ \vect{v}^2   + \tau^2(\vect{v}\gvect{\omega})(\vect{v}\gvect{\omega}) + \tau \vect{v} (\vect{v} \times \gvect{\omega}) \right] \boldsymbol{\mathcal{E}}
\end{eqnarray}
where we have taken into account that the term containing $f_0$ vanishes due to spherical symmetry, and 2 is the spin factor. Momentum integration yields
\begin{eqnarray}
  \nonumber
  \label{eq:19}
  2\int d\vect{p}\left(-\pdif{f_0}{E} \right)  = \frac{\mu p_F}{\pi^2} \\
  \times \left\{1 + \frac{1}{3}\left(\frac{\pi T}{\mu} \right)^2  +  O\left[\left(\frac{T}{\mu}\right)^4 \right] + ... \right\}\\
  \nonumber
  =  \nu
\end{eqnarray}
This expression for the density of states was already introduced in (\ref{eq:7}). Finally we obtain the EC tensor(constant MF is directed along the z-axis)
\begin{equation}
  \label{eq:20}
  \sigma_{xx} = \sigma_{yy} = \frac{1}{3}e^2v^2\nu \frac{\tau}{1 + \Omega^2\tau^2} = \sigma_0\frac{1}{1 +\Omega^2\tau^2}
\end{equation}
\begin{equation}
  \label{eq:21}
  \sigma_{zz} = \sigma_0
\end{equation}
\begin{equation}
  \label{eq:22}
  \sigma_{xy} = \sigma_0 \frac{\tau \Omega}{1 + \Omega^2 \tau^2} = - \sigma_{yx}
\end{equation}
where $\sigma_0$ is Drude EC given by (\ref{eq:10}) at $\omega = q = 0$. In line with symmetry requirements EC along MF remains unchanged \cite{14}. Summarizing, we point out the distinctions between the above results and the standard ones \cite{13,14}. The density of states (\ref{eq:19}) corresponds to relativistic kinematics and the condition $\mu \gg T$. The damping of EC by MF due to factor the $(1 + \Omega^2\tau^2)^{-1}$ is for $\mu \gg m$ much slower than by a similar factor with $\Omega' = eB/m$ in the usual case.

\section{Paraconductivity}
\label{sec:4}

Next we turn to the quantum contribution to EC. In superconductors such phenomena have been intensively studied for more than three decades \cite{4}. Precritical phenomena in quark matter and fluctuation conductivity(FC), or paraconductivity, have been discussed in \cite{16} without deriving explicit equations. 

First we want to point out an important difference between fluctuating phenomena in superconductors \cite{4} and in high density quark matter \cite{3,17}. Attraction between electrons in a superconductor is due to the interaction with the lattice(phonon mechanism). The characteristic energy scale of this interaction is the Debye frequency $\omega_D \simeq 1/2(m/M)^{1/2} \ Ry \sim 10^{-2} \ eV$, where $m$ and $M$ are the electron and ion masses. On the other hand, the typical Fermi energy is $E_F = p_F^2/2m \simeq (3\pi^2n)^{2/3}/2m \sim 2 \ eV$. Thus $\omega_D \ll E_F$ and the interaction in the BCS regime is concentrated within a very thin layer around the Fermi surface. On the other hand, the role of $\omega_D$ in color superconductor is played by the momentum cutoff $\Lambda \sim 0.7 \ GeV$ \cite{1,2} while in high density and moderate temperature regime $E_F \simeq \mu \sim 0.4 \ GeV$. Thus the scale hierarchy is opposite to that in BCS. Another difference concerns the size of pairs forming the condensate. In BCS this is the coherence length $\xi$ which is macroscopic, i.e., much larger than the interatomic distance $\xi \sim 10^{-4} \ cm$, $a \sim 10^{-8} \ cm$. The BCS dimensionless parameter is $k_F\xi \geq 10^3$. In color superconductor the same quantity is $k_F\xi \sim 2$ \cite{17}, where $\xi$ is the size of the diquark pair, $k_F \simeq \mu$. The fluctuation contribution to the physical quantities is characterized by Ginzburg-Levanyuk number $Gi$ which for the quark matter may be estimated as 
\begin{equation}
  \label{eq:23}
  Gi \simeq \left(\frac{T_c}{\mu}\right)^4 \sim 10^{-2} - 10^{-3}
\end{equation}
We remind that for the BCS superconductor $Gi \sim 10^{-12} - 10^{-14}$.

Two conclusions may be drawn from the above considerations. Fluctuation region in quark matter is much wider than in BCS superconductors. Quantum contribution to EC may be substantial contrary to the situation in solid state physics where it is called ``quantum correction''. 

The theoretical study of fluctuation conductivity dates from the seminal paper \cite{19} and since then the subject has been intensively studied in the framework of both conventional and high-temperature superconductivity theories \cite{4}. In strong coupling regime paraconductivity has been discussed in \cite{20}. 

\begin{figure*}
  \includegraphics[width=0.8\textwidth]{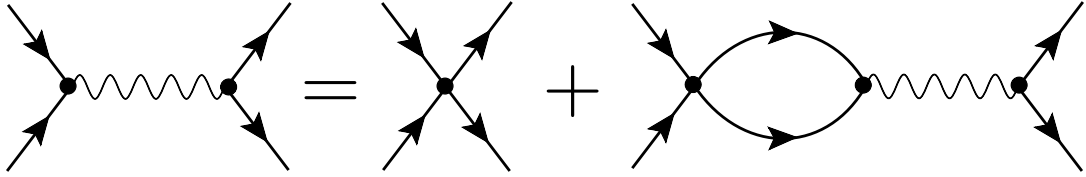}
  \caption{Dyson equation for FP (wavy line)}
  \label{fig:2}
\end{figure*}

At temperature higher than $T_c$ quark pairing is energetically unfavourable. As soon as the temperature approaches $T_c$ from above the number of fluctuation quark pairs increases. Such pairs are described by the fluctuation propagator(FP) \cite{4,17,19,21} represented by the Feynman diagram in Fig.\ref{fig:2}.
Analytically  Dyson equation for the FP $L(\vect{q}, \omega_k)$ reads
\begin{equation}
  \label{eq:24}
  L(\vect{q}, \omega_k) = \frac{1}{-\frac{1}{g} + \Pi(\vect{q}, \omega_k)}
\end{equation}
Here $g$ is the coupling constant with dimension $m^{-2}$, $\Pi(\vect{q}, \omega_k)$ is the polarization operator. If for a moment we ignore that quarks undergo chaotic scattering, the polarization operator would have the form
\begin{widetext}
  \begin{equation}
    \label{eq:25}
    \Pi(\vect{q}, \omega_k) = T \sum_{\varepsilon_n} \int \frac{d\vect{p}}{(2\pi)^3} G(-\vect{p},-\varepsilon_n)G(\vect{p} + \vect{q}, \varepsilon_n + \omega_k) = T\sum_{\mathcal{E}_n}P(\vect{q}, \varepsilon_n + \omega_k, -\varepsilon_n)
  \end{equation}
\end{widetext}
Note the sign difference between the arguments of the Green's functions in (\ref{eq:1}) and (\ref{eq:25}) and opposite directions of the corresponding arrow in Figs.\ref{fig:1} and \ref{fig:2}. This is because the FP corresponds to the Cooper channel. In the vicinity of $T_c$ fluctuations are dominated by long-wave modes. Therefore $P(\vect{q}, \varepsilon_n + \omega_k, -\varepsilon_n)$ can be expanded as \cite{4,3,21,22}
\begin{equation}
  \label{eq:26}
  P(\vect{q}, \varepsilon_n + \omega_k, -\varepsilon_n) = A(\vect{q} = 0, \omega_k, \varepsilon_n) + B(\omega_k, \varepsilon_n)\vect{q}^2
\end{equation}
Calculation of $A$ and $B$ is straightforward and were presented in \cite{3,22}. One has 
\begin{equation}
  \label{eq:27}
  P(\vect{q}, \varepsilon_n + \omega_k, -\varepsilon_n) = \frac{2\pi\nu_0}{|2\varepsilon_n + \omega_k|} \left[1 - \frac{v^2 \vect{q}^2}{3}\frac{1}{|2\varepsilon_n + \omega_k|^2} \right]
\end{equation}
Here $\nu_0 = \mu p_F/\pi^2$ is the relativistic density of states at the Fermi level. As we shall see below, the first term in (\ref{eq:26}-\ref{eq:27}) gives after summation over $\varepsilon_n$ the Cooper logarithm and the second one corresponds to the long-wave fluctuations. 

\begin{figure*}
  \includegraphics[width=0.8\textwidth]{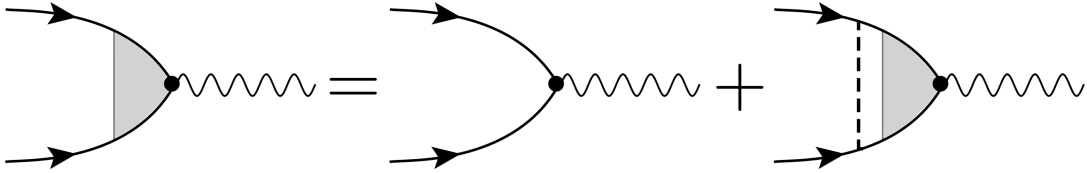}
  \caption{The equation for the vertex part in the ladder approximation. The dashed line corresponds to the random correlator.}
  \label{fig:3}
\end{figure*}

At the next step we have to include the random scattering and the corresponding momentum relaxation time parameter $\tau$. The averaging of the Green's functions over the disorder amounts to the replacement of $\varepsilon_n$ in $P(\vect{q}, \varepsilon_n + \omega_k, -\varepsilon_n)$ by $\tilde{\varepsilon}_n = \varepsilon_n + \frac{1}{2\tau}\mathrm{sgn}\varepsilon_n$. The averaging procedure includes also the renormalization of the vertex function which is represented graphically in Fig.\ref{fig:3}.
As a result the polarization operator takes the form 
\begin{equation}
  \label{eq:28}
  \Pi(\vect{q}, \omega_k) = T \sum_{\mathcal{E}_n} \frac{1}{P^{-1}(\vect{q}, \varepsilon_n + \omega_k, -\varepsilon_n) - \langle U^2 \rangle }
\end{equation}
Here $U(r)$ is a potential responsible for the quark momentum relaxation. If we adopt the standard assumption that it satisfies the Gauss law, then after the averaging over the volume we have $2\pi \nu_0 \langle U^2 \rangle = \tau^{-1}$ \cite{4,21}. Equation (\ref{eq:28}) may be formulated in terms of the vertex renormalization correction $\lambda$ as
\begin{equation}
  \label{eq:29ins}
  \Pi(\vect{q}, \omega_k) = T \sum_{\varepsilon_n}\lambda(\vect{q}, \varepsilon_n + \omega_k, -\varepsilon_n) P(\vect{q}, \varepsilon_n + \omega_k, -\varepsilon_n), 
\end{equation}
\begin{equation}
  \label{eq:30ins}
  \lambda^{-1}(\vect{q},\varepsilon_1, \varepsilon_2) = 1 - \frac{1}{2\pi \nu_0 \tau} P(\vect{q},\tilde{\varepsilon}_1, \tilde{\varepsilon}_2).
\end{equation}
In particular, at $\vect{q} \rightarrow 0$, $\omega_k \rightarrow 0$ from (\ref{eq:27}) we obtain 
\begin{equation}
  \label{eq:31ins}
  \lambda(\vect{q}, \varepsilon_n + \omega_k, -\varepsilon_n) \rightarrow \frac{|\tilde{\varepsilon}_n|}{|\varepsilon_n|}.
\end{equation}

 Performing the summation over $\varepsilon_n$ in (\ref{eq:28}) we obtain
\begin{eqnarray}
  \nonumber
  \label{eq:29}
  \Pi(\vect{q}, \omega_k)  = \nu_0 \left[\ln \frac{\Lambda}{2\pi T}  \right.\\
  \left. - \psi \left(\frac{1}{2} + \frac{|\omega_k|}{4\pi T} \right)  - \frac{\pi}{8T} \mathcal{D} \vect{q}^2 \right]
\end{eqnarray}
Here $\psi(x)$ is the logarithmic derivative of the $\Gamma$-function, $\Lambda$ is the high-frequency cut-off, $\Lambda \gg |\omega_k|$, $\mathcal{D}$ is the diffusion coefficient which can be formally introduced by the equation
\begin{equation}
  \label{eq:30}
  \mathcal{D} = \frac{7\zeta(3)v^2}{6\pi^3T} \chi \left(\frac{1}{2\pi T \tau} \right)
\end{equation}
\begin{equation}
  \label{eq:31}
  \chi(x) = \frac{8}{7\zeta(3)}\sum_{n=0}^{\infty} \frac{1}{(2n+1)^2(2n + 1 + x)}
\end{equation}
In the limiting cases (\ref{eq:30}-\ref{eq:31}) yield

\begin{equation}
  \label{eq:32}
 \mathcal{D} = 
    \begin{cases}
       \frac{v^2\tau}{3}, \ T\tau \ll 1 \\
       \frac{v^2}{6T}, \ T\tau \gg 1
    \end{cases} 
\end{equation}
As we shall discuss below, the quark matter in the precritical regime corresponds to $T\tau \leq 1$. Assuming that $|\omega_k| \ll 4\pi T$ we expand the second term in (\ref{eq:29}) and then replace $\Lambda$ by the critical temperature using the Thouless condition
\begin{eqnarray}
  \nonumber
  \label{eq:33}
  \ln \frac{\Lambda}{2\pi T} - \psi \left(\frac{1}{2} \right) \\
  \nonumber
  = \left(\ln \frac{\Lambda}{2\pi T_c} - \psi \left(\frac{1}{2} \right) \right) - \mathcal{E}\\
  = \frac{1}{g\nu_0} - \mathcal{E}.
\end{eqnarray}
Here $\mathcal{E} = \ln \frac{T}{T_c} \simeq \frac{T - T_c}{T_c}$. Finally we can write the following expression for the FP (\ref{eq:24}) ($\omega = i\omega_k$)
\begin{equation}
  \label{eq:34}
  L(\vect{q}, \omega) = -\frac{1}{\nu_0}\frac{1}{\mathcal{E} + \frac{\pi}{8T}\left(-i\omega + \mathcal{D}\vect{q}^2 \right)}
\end{equation}
Two remarks are appropriate at this point. First, the only difference of the FP (\ref{eq:34}) from its nonrelativistic counterpart \cite{4,21} amounts to a replacement of the nonrelativistic density of states by the relativistic one $\nu_0$. Second, as it was shown in \cite{20}, the derivation of the FP in the strong coupling regime does not alter the final result (\ref{eq:34}).

\begin{figure}[!h]
  \includegraphics[width=0.48\textwidth]{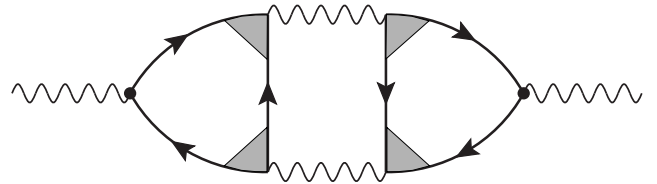}
  \caption{Feynman diagram representing the AL term. Internal wavy lines are FP-s, solid lines are quark propagators, the rectangles are the renormalized vertices $\lambda$.}
  \label{fig:4}
\end{figure}

With the FP at our disposal we can evaluate the Aslamazov-Larkin(AL) (paraconductivity) contribution to EC in the fluctuation region. Based on the experience gained in condensed matter physics \cite{4} we assume that it is of major importance among other quantum fluctuation effects. The corresponding Feynman diagram is shown in Fig.\ref{fig:4}
The EC can be written as
\begin{equation}
  \label{eq:35}
  \sigma_{lm}(AL) = \frac{1}{\omega_k} Q_{lm}(\omega_k)
\end{equation}
where $Q_{lm}(\omega_k)$ is the retarded electromagnetic response operator and for the AL contribution it is given by the Feynman diagram shown in Fig.\ref{fig:4} and reads \cite{19,4,21}
\begin{widetext}
\begin{equation}
  \label{eq:36}
  Q_{lm}(\omega_k) = e^2T \sum_{\Omega_j} \int \frac{d\vect{q}}{(2\pi)^3} B_l(\vect{q}, \Omega_j, \omega_k) L(\vect{q},\Omega_j)B_m(\vect{q}, \Omega_j, \omega_k)L(\vect{q}, \Omega_j +\omega_k),
\end{equation}
where $L$ are FP-s, and $B_{l,m}$ correspond to the three Green's functions blocks
\begin{equation}
  \label{eq:37}
  \vect{B}(\vect{q}, \Omega_j, \omega_k) = eT \sum_{\varepsilon_n} \lambda(\vect{q}, \varepsilon_n + \omega_k, \Omega_j - \varepsilon_n) \lambda(\vect{q}, \varepsilon_n, \Omega_j - \varepsilon_n) \int \frac{d \vect{p}}{(2\pi)^3} \mathrm{Tr} \left[ \gvect{\gamma} G(\vect{p},\tilde{\varepsilon}_n)G(\vect{p},\tilde{\varepsilon}_n + \omega_k)G(\vect{q} - \vect{p}, \Omega_j - \tilde{\varepsilon}_n)  \right].
\end{equation}

\end{widetext}
The propagators entering into (\ref{eq:37}) may be decomposed into contributions from quarks and antiquarks
\begin{equation}
  \label{eq:41ins}
  G(\vect{p},\tilde{\varepsilon}_n) = - \frac{1}{2E} \left[ \frac{\gamma_0E - \gvect{\gamma} \vect{p} + m}{(E - \mu) - i\tilde{\varepsilon}_n} - \frac{\gamma_0E + \gvect{\gamma}\vect{p} - m}{(E + \mu) + i\tilde{\varepsilon}_n} \right].
\end{equation}
We are considering the high density regime when the system is dopped with excess quarks. At high $\mu$ the antiquark part of the Green's function is suppressed by the factor $(E + \mu)$ is the denominator and will be dropped. We plan to investigate the accuracy of this approximation as a function of the $\mu$ in a forthcoming study. Close to the Fermi surface the physical quantities depend on the variable $\xi = E - \mu = \sqrt{\vect{p}^2 + m^2} - \mu$. The high density Green's function takes the form
\begin{equation}
  \label{eq:42ins}
  G(\vect{p},\tilde{\varepsilon}_n) = -\frac{1}{2E} \frac{\gamma_0E - \gvect{\gamma}\vect{p} + m}{\xi - i\tilde{\varepsilon}_n}.
\end{equation}
In the vicinity of $T_c$ the FP (\ref{eq:34}) has a pole structure. The dependence of $L(\vect{q}, \Omega_j)$ and $L(\vect{q}, \Omega_j - \omega_k)$ on $\Omega_j$ and $\omega_k$ is much stronger than the dependence of the Green's functions on the same quantities. We shall keep in the propagators entering into $\vect{B}(\vect{q},\Omega_j,\omega_k)$ only the dependence on the fermionic frequencies $\tilde{\varepsilon}_n$ and calculate $\vect{B}(\vect{q}, \Omega_j = \omega_k = 0)$. 

Expanding $G(\vect{q} - \vect{p}, - \tilde{\varepsilon}_n)$ at $\vect{q} \rightarrow 0$ one has
\begin{eqnarray}
  \nonumber
  \label{eq:43ins}
  G(\vect{q} - \vect{p}, - \tilde{\varepsilon}_n) \simeq G(-\vect{p}, -\tilde{\varepsilon}_n) + \vect{q} \pdif{}{\vect{p}} G(-\vect{p},-\tilde{\varepsilon}_n) \simeq\\
G(-\vect{p}, -\tilde{\varepsilon}_n) + \frac{(\vect{q} \vect{p})}{\mu} \pdif{}{\xi}G(-\vect{p}, -\tilde{\varepsilon}_n).
\end{eqnarray}
Then we have
\begin{eqnarray}
  \nonumber
  \label{eq:44ins}
  \mathrm{Tr} \left[\gvect{\gamma}G(\vect{p},\tilde{\varepsilon}_n)G(\vect{p},\tilde{\varepsilon}_n)G(\vect{q} - \vect{p}, -\tilde{\varepsilon}_n) \right] \simeq \\
  \frac{2\vect{p}}{\mu} \left[\frac{1}{(\xi - i \tilde{\varepsilon}_n)^2(\xi + i\tilde{\varepsilon}_n)} + \frac{(\vect{q}\vect{p})}{\mu}\frac{1}{(\xi^2 + \tilde{\varepsilon}_n^2)^2} \right].
\end{eqnarray}
Angular integration over $\Omega_{\vect{p}}$ kills the first term. The second term yields 
\begin{eqnarray}
  \nonumber
  \label{eq:45ins}
  B_l(\vect{q}) = -\frac{e\nu_0}{3} \frac{\vect{p}^2}{\mu^2} q_l T \sum_{\tilde{\varepsilon}_n}\frac{|\tilde{\varepsilon}_n|^2}{|\varepsilon_n|^2} \int^{\infty}_{-\infty} \frac{d \xi}{(\xi^2 + \tilde{\varepsilon}_n^2)^2} = \\
- \frac{e\pi \nu_0}{T_c}\frac{v^2 \tau}{12} q_l = -\tilde{B} q_l,
\end{eqnarray}
where $v^2 = p^2/\mu^2$ is the quark velocity at the Fermi surface. The vertex corrections were taken in the form (\ref{eq:31ins}) since the $\vect{q}$-dependence is essential only at $\mathcal{D}\vect{q}^2 \sim \mathcal{E} \sim T$. The summation in (\ref{eq:45ins}) has been performed using (\ref{eq:30}) - (\ref{eq:32}). Returning to (\ref{eq:36}), we write
\begin{equation}
  \label{eq:46ins}
  Q_{lm}(\omega_k) = \tilde{B}^2T \sum_{\Omega_j} \int \frac{d\vect{q}}{(2\pi)^3} q_l q_m L(\vect{q},\Omega_j)L(\vect{q},\Omega_j + \omega_k).
\end{equation}
To evaluate the sum in (\ref{eq:46ins}) we can use the technique of replacing the summation by contour integration \cite{26,27}. At the first step this leads to the following result for the absolute value of the response operator 
\begin{eqnarray}
  \nonumber
  \label{eq:47ins}
  Q(\omega) = \frac{\tilde{B}^2}{6\pi} \int \frac{d\vect{q}}{(2\pi)^3} \vect{q}^2 \int_{-\infty}^{\infty} dz \coth{\frac{z}{2T}} \\
\left[L^R(\vect{q}, -iz-i\omega) + L^A(\vect{q},-iz +i\omega) \right] \mathrm{Im} L^R(\vect{q},-iz),
\end{eqnarray}
where $z = i\Omega_j$, $\omega = i\omega_k$, $L^R$ and $L^A$ are the retarded and advanced FP-s. The next step is to expand the integrand in powers of $\omega$ and to subtract the zeroth order term which would lead to Meissner effect above $T_c$. This may be regarded as imposing the Ward identity by brute force. The problem is a subtle one \cite{28,29}. To take care to satisfy the Ward identity one has to consider the sum of the AL and Maki-Thompson \cite{30,31} diagrams. Nonetheless, it is almost always to take into account only the AL contribution to fit experimental data above the transition in ordinary superconductors and HTS compounds \cite{4}. 

Keeping in (\ref{eq:47ins}) the term proportional to $\omega$ and integrating by parts, one has 
\begin{eqnarray}
  \nonumber
  \label{eq:48ins}
  Q(\omega) = -i\omega \tilde{B}^2 \frac{T}{3\pi} \int \frac{d\vect{q}}{(2\pi)^3} \vect{q}^2 \int_{-\infty}^{\infty} dz \frac{(\mathrm{Im} L^R)^2}{z^2} =\\
    -i\omega \frac{\pi \tilde{B}^2}{12 \nu_0^2} \int \frac{d\vect{q}}{(2\pi)^3} \frac{\vect{q}^2}{\left(\mathcal{E} + \frac{\pi}{8T_c}\mathcal{D} \vect{q}^2 \right)^3}.
\end{eqnarray}
The final result for the conductivity reads
\begin{equation}
  \label{eq:49ins}
  \sigma(AL) = \frac{e^2}{16}(\alpha \mathcal{E})^{-1/2},
\end{equation}
where 
\begin{equation}
  \label{eq:50ins}
  \alpha = \frac{\pi v l}{24 T_c} = \frac{\pi}{8T_c} \mathcal{D},
\end{equation}

where $\mathcal{D}$ is the diffusion operator equal to $\mathcal{D} = \frac{1}{3}vl$ in the "dirty" limit.(see (\ref{eq:32})). The important point that in the fluctuation region the parameter for the quark system is large 
\begin{equation}
  \label{eq:40}
  \mathcal{E} \simeq \frac{T - T_c}{T_c} \simeq Gi \sim 10^{-2}
\end{equation}
The long-wave fluctuation picture which led to the result (\ref{eq:50ins}) becomes inadequate very close to $T_c$ where different physical mechanisms should be included. Now we have to compare the relative contributions of the Drude and AL EC and briefly compare our results with the numerious calculations of EC presented in literature. 

\section{Conclusions and a Look Ahead}
\label{sec:5}

Our first important conclusion is that in the precritical region of high density and moderate temperature the fluctuation EC(paraconductivity) not only reaches the value of the Drude one, but can greatly exceed it. To make this statement clear-cut we combine Eqs.(\ref{eq:10}) and (\ref{eq:38}) at $|\vect{q}|  = \omega = 0$ in the following form
\begin{eqnarray}
  \nonumber
  \label{eq:41}
  \sigma = \sigma(Drude) + \sigma(AL) \\
  = \frac{e^2}{3}\nu_0vl \left(1 + \frac{3}{8 \sqrt{\mathcal{E}}\nu_0vl}\sqrt{\frac{6T_c}{\pi vl}} \right)
\end{eqnarray}
where $l = v\tau$. As it was already noted, our knowledge of the QCD phase diagram in the high density region is far from being settled \cite{23}. The hope is that the future experiments at FAIR and NICA will bring a much more reliable picture. So far we can take $\mu = 300 \ MeV$, $T_c = 80 \ MeV$ as a reasonable set of parameters in the precritical region \cite{23}. For simplicity we take $v = 1$, so that $\nu_0 = \mu^2/\pi^2$, $l = \tau$. The most important parameter is $\tau$ for which to our best knowledge a reliable determination is lacking. For example, in \cite{24} it varies within the interval $0.3 \ fm \ < \tau < \ 0.9 \ fm$. As an estimate we take $\tau = 0.5 \ fm$. Equation (\ref{eq:41})takes the form 
\begin{eqnarray}
  \nonumber
  \label{eq:42}
  \sigma = \sigma(Drude)\left(1 + \frac{5}{\sqrt{\mathcal{E}}} \frac{\sqrt{T_c \tau}}{(\mu \tau)^2} \right) \\
   \simeq \sigma(Drude)\left(1 + \frac{4}{\sqrt{Gi}} \right)
\end{eqnarray}
We see that for $Gi \sim 10^{-2}$ paraconductivity dominates over the Drude one. In condensed matter physics the situation when paraconductivity becomes equal to the classical one is possibly realized in the two-dimensional samples \cite{4}. 

Another conclusion is the stabilization of the Drude EC in MF up to the MF values $\left(eB/\mu \right)\tau \sim 1$ - see Eq.(\ref{eq:20}). For our choice of parameters the corresponding MF value is $eB \sim 0.6 \cdot 10^{19} \ G$. The problem of the  paraconductivity dependence on the MF will be the subject of the forthcoming dedicated publication.

Finally, let us present for the orientation purposes the values of the Drude and AL EC for the above set of parameters. We have $e^2 = C_{EM} \sim 0.051$(see (\ref{eq:2})). Then 
\begin{equation}
  \label{eq:43}
  \sigma(Drude) \sim 0.002 \ fm^{-1},\ \sigma(AL) \sim 0.08 \ fm^{-1}
\end{equation}
As already mentioned, the EC in this domain of the QCD phase diagram has not been calculated before. Numerous results at $\mu = 0$ and different values of $T$ \cite{24,15} differ from each other by an order of magnitude. As an example we quote Ref.\cite{24} according to which $\sigma \sim (0.02 \sim 0.15) \ fm^{-1}$ for $T \sim (0 \sim 400) \ MeV$ and $\mu = 0$. We note in passing that our value of $\sigma(Drude)$ is about two orders of magnitude larger than the EC of Cu at room temperature, while $\sigma(AL)$ exceeds it by a factor 5000.

Our study of the EC near the transition line at high density allows to suggest that other transport coefficients - shear viscosity and thermal conductivity will be dominated by the fluctuation effects as well. Another important physical observable which may display a spectacular behaviour in the fluctuation domain is the lepton-pair production rate which is determined by the current-current correlator. 

B.K. is indebted to A. Varlamov for illuminating discussions. M.A. is grateful to the Dynasty Foundation. The research was supported by RFBR grant 1402-00395.

\end{document}